\documentclass[aps,pra,twocolumn,amsmath,amssymb,showpacs,10pt,amsfonts,longbibliography]{revtex4-1}
\usepackage{epsfig}
\usepackage{graphics}
\usepackage{dcolumn}
\usepackage{bm}
\usepackage{epstopdf}
\usepackage{color}
\usepackage{amsmath}
\usepackage{psfrag}
\usepackage{subfigure}

\begin{document}
\title{Inertial particles distribute in turbulence as Poissonian points with random intensity inducing clustering and supervoiding}
\author{Lukas Schmidt$^{1}$}
\author{Itzhak Fouxon$^{1,2}$}
\author{Markus Holzner$^1$}
\affiliation{$^1$ ETH Zurich, Wolfgang-Pauli-Strasse 15, 8093 Zurich, Switzerland} \affiliation{$^2$ Department of Computational Science and Engineering, 
Yonsei University, Seoul 120-749, South Korea} 

\begin{abstract}
This work considers the distribution of inertial particles in turbulence using the point-particle approximation. We demonstrate that the random point process formed by the positions of particles in space is a Poisson point process with log-normal random intensity (the so-called "log Gaussian Cox process" or LGCP). The probability of having a finite number of particles in a small volume is given in terms of the characteristic function of a log-normal distribution. Corrections due to discreteness of the number of particles to the previously derived statistics of particle concentration in the continuum limit are provided. These are relevant for dealing with experimental or numerical data. The probability of having regions without particles, i.e. voids, is larger for inertial particles than for tracer particles where voids are distributed according to Poisson processes. 
The ratio of the typical void size to the average concentration raised to the power of $-1/3$ is of order one in the limit of zero inertia at a fixed total number of particles. However at fixed inertia the ratio diverges in the limit of infinite number of particles. Thus voids are very sensitive to inertia. Further, the probability of having large voids decays only log-normally with size. This shows that particles cluster, leaving voids behind. Remarkably, at scales where there is no clustering there can still be an increase of the void probability so that turbulent voiding is stronger than clustering. The demonstrated double stochasticity (Poisson with random intensity) of the distribution originates in the two-step formation of fluctuations. First, turbulence brings the particles randomly close together which happens with Poisson-type probability. Then, turbulence compresses the particles' volume in the observation volume. We confirm the theory of the statistics of the number of particles in small volumes by numerical observations of inertial particle motion in a chaotic $ABC$ flow. The improved understanding of clustering processes can be applied to predict the long-time survival probability of reacting particles. Our work implies that the particle distribution in weakly compressible flow with finite time correlations is a LGCP, independently of the details of the flow statistics.      
\end{abstract}\maketitle

\section{Introduction}\label{sec:intro}

Studies of distributions of dilute suspensions of inertial particles in turbulence have received increasing attention recently \cite{Monchaux,Bala,reade,boffetta1,BEC_BIFERALE_LANOTTE_SCAGLIARINI_TOSCHI_2010,Falkovich,Stefano,fouxonlee,xu2008,Calzavarini,Collinstwo}. The dilute suspension limit of the highly challenging problem of multiphase turbulent flows provides the reference point for studies of dense solutions. Dilute particle distribution implies that the change of the fluid flow by particles is negligible so the particle transport by the fluid takes place in a given undisturbed turbulent flow. Nonetheless particles are not perfect tracers because their inertia leads to deviations from the flow trajectories. This creates inhomogeneous spatial distributions in regions where fluctuations of particle concentration can be large. The solution of the problem for weakly inertial particles revealed divergent root mean square fluctuation of concentration in the continuum approximation \cite{fouxon1}. This manifests preferential concentration or clustering of particles. Applying the usual procedure of finding concentration as the number of particles in a small volume divided by the volume and taking zero volume as limit does not result in a well-defined outcome. The number of particles in a small ball, scales with the ball's radius but the scaling exponent changes from point to point manifesting a multifractal distribution of particles in space. This is in contrast with smooth distributions where the number of particles scales as the third power of the ball's radius. It is thus a reasonable approach to study the statistics of the number of particles directly in the small volume with the aim to determine the counterpart of the Poisson distribution that holds for this problem in ideal gases. To the best of our knowledge this has not been done so far. 
Previous studies used the continuum approximation that works when the average number of particles in relevant volumes is large. This is not necessarily always the case. In one of the main applications of inertial particle clustering in turbulence - distribution of water droplets transported by air turbulence in warm clouds - the number of droplets per viscous scale of turbulence does often not exceed one \cite{review}. It is however at this scale that preferential concentration of particles happens. Thus discreteness of matter is relevant. It is relevant also for providing the correspondence between theory and experiments where particles are discrete necessarily. Furthermore in numerical simulations it is often not feasible to simulate a large number of particles below the viscous scale which is the smallest scale of turbulence and generally several orders of magnitude smaller than the energy-containing eddies of the flow \cite{frisch}. 

In this work we provide the distribution of the number of inertial particles in a small volume of turbulent flows. We find that this distribution samples a Poisson distribution with random intensity. This provides a new way of thinking of the clustering effect of inertial particles turbulence since intensity fluctuations increase the probability of particles being close together. Our prediction holds for arbitrary homogeneous, chaotic flow. We confirm the predictions numerically using a synthetic, random $ABC$ flow. 

Our study uses the traditional framework where turbulent drag of the particles is linear in the difference of the velocities of the particles and the local flow. Considering a finite number of particles, the random transport of particles by turbulence defines the random point processes in space where points represent the positions of the spherical particles' centers. The statistics of this random process has not yet been studied directly. Instead so far the continuum approximation for the particle concentration was used. This approximation results in the radial distribution function, which is the probability of finding pairs of particles with fixed separation, which has been the main object of previous studies. 

Another object of physical interest studied recently, is the probability of holes of particles or hollows \cite{boffetta_delillo,goto}. This is the probability of having no particles inside an arbitrary given volume for which a discrete formulation is most reasonable. Holes can play a decisive role in long-term survival in the case where chemical reactions occur, see \cite{zeldovichovchinnikov}. Further the probability of holes plays a central role in the study of random point processes where it is called void probability. The reason is that the void probability determines the statistics of the point process completely, see e. g. \cite{moller2} and references therein. We provide the void probability demonstrating that it is larger than for Poissonian random point processes. This is the manifestation of clustering where particles accumulate in certain regions leaving voids behind. 

In this work we study the range of parameters where particles' trajectories are unique not only in the phase space (where Newton's law holds) but also in the physical space. In this case the particles' motion obeys a smooth spatial flow. That flow is the flow of particles which is different from the flow of the fluid. For a long time the particles' flow description was known to hold for weakly inertial particles \cite{maxey}. It was found recently that the flow description can be introduced in the case of strong gravity as well \cite{fouxonlee}. The difference between the particles' and the fluid flow is caused by the combined effect of inertia and gravity that separate the particles from the local fluid. The existence of the flow of particles is used for the introduction of the continuum description of the particle concentration using the continuity equation. The continuum theory uses the weak compressibility of the particle flow. This allows using the general theory developed in \cite{fouxon1} for describing the particle statistics. The concentration of particles transported by weakly compressible flow obeys log-normal statistics. The theory predicts that there is a random attractor in space to which the inertial particles' trajectories converge after transients \cite{FFS}.  Particle statistics are determined completely by the pair-correlation function for which the simple closed form holds.
An initially uniform or continuous particle concentration will become supported in a singular multifractal structure in space after transients \cite{fouxon1}. This multifractal is by itself time-dependent and changes continually over time keeping its statistical, space-averaged properties constant. 

In the following, we analyze how to understand these predictions in the framework of discrete particles. The limiting case of only one inertial particle in the system inserted in the system at time $t=0$ is considered here. At the same time $t=0$, there is a multifractal set in the flow. This multifractal is where the particles that would be hypothetically inserted in the flow in the remote past would find themselves at $t=0$. This mathematical set evolves in time. The inserted particle will "find" this multifractal with time and "stick" to it never leaving it again (except for occasional deviations with very small probability). Inserting a finite number $N$ of particles at $t=0$ would cause all particles to find their place on the same multifractal, filling it in some sparse way. Only for the limit $N\to\infty$ the particles would form the spatial pattern manifesting this multifractal structure. 

The continuum theory does provide basic predictions on the discrete particles. Here, we use the pair-correlation function of the particle concentration that provides the probability of finding another particle at a fixed distance from the given particle. We use this prediction for finding the collision kernel of particles relevant whenever two-body collisions occur in the system with its basic application in the rain formation problem. 
Practical applications where the presented theoretical considerations are of interest are studies investigating the statistics of the number of aerosol particles caught by the particle detector that since recently has been accessible in experiments (e.g.\cite{Siebert,toprak,ahmad}). 

This work is structured as follows, first we discuss the current available continuum predictions on inertial particle statistics in turbulent flows. Then we derive the statistics of the number of particles in a small volume, considering its discreteness (Section \ref{sec:theory}). Additionally, we also derive predictions for the probability of finding voids of particles and investigate the size of these voids. In Section\,\ref{sec:rdf} we demonstrate that continuum and discrete predictions coincide using the radial distribution function. Subsequently, we confirm the theoretical derivations of the number of particles in a small volume numerically with particles advected in a chaotic $ABC$ flow (Section \ref{sec:numerics}).

\section{Framework of the study and continuum results}\label{sec:stateoftheart}

In this Section we introduce the equations of particle motion and summarize the statistics of the particle concentration obtained previously using the continuum approximation.

We consider the motion of particles in an incompressible turbulent flow $\bm u(t, \bm x)$. The particles concentration is assumed to be small so both the particles' interactions between themselves and their reaction to turbulence can be neglected (so called one-way coupling). Thus we can concentrate on the motion of one particle. Particles are considered spherical with radius $a$ much smaller than the smallest scale of the spatial variations of turbulence, $\eta$. Here, $\eta=(\nu^3/\epsilon)^{1/4}$ is the viscous Kolmogorov scale where $\epsilon$ is the average rate of energy dissipation per unit volume of the fluid and $\nu$ is the kinematic viscosity. If the Reynolds number $Re_p$ of the flow perturbation caused by the particle is small then the linear law of friction can be used,  
\begin{eqnarray}&&
\frac{d\bm x}{dt}=\bm v,\ \ \frac{d\bm v}{dt}=-\frac{\bm v-\bm u[t, \bm x(t)]}{\tau}+\bm g,\label{lw}
\end{eqnarray}
where $\bm x$, $\bm v$ are the particle's coordinate and velocity respectively. Here $\bm g$ is the gravitational acceleration and $\tau=2\rho_p a^2/[9 \nu \rho ]$ is the Stokes time, where $\rho_p$, $\rho$ are the particle and fluid densities respectively. This linear law can be used for $Re_p\lesssim 1$ provided that the $Re_p-$ dependence is introduced in $\tau$ that becomes relevant at $Re_p\sim 1$. For instance for water droplets in clouds the linear friction law can be used up to a droplet size of $50$ microns where $Re_p\sim 1$, see \cite{fouxonlee,review,ayala}. Below, $\tau$ is used as the effective velocity relaxation time in Eq.~(\ref{lw}) including the $Re_p-$dependence.

In the case where gravity can be neglected and the Stokes number $St=\tau\sqrt{\epsilon/\nu}$ is small after transients the velocity of the particle is determined uniquely by its position, 
\begin{eqnarray}&&
\bm v=\bm u[t, \bm x(t)]-\tau\left[\partial_t\bm u+(\bm u\cdot\nabla)\bm u\right][t, \bm x(t)].\label{eq:particlevelocity}
\end{eqnarray}
Thus the flow of particles $\bm v(t, \bm x)=\bm u-\tau\left[\partial_t\bm u+(\bm u\cdot\nabla)\bm u\right]$ can be introduced so that if there is a particle at $\bm x(t)$ then its velocity is $\bm v(t, \bm x)$ at that point,  
\begin{eqnarray}&&
\frac{d\bm x}{dt}=\bm v[t, \bm x(t)],\ \ \nabla\cdot\bm v\neq 0,\label{flow}
\end{eqnarray}
where we stress that the particles' flow, in contrast with the underlying turbulent flow of the fluid, has finite compressibility, $\nabla\cdot\bm v=-\tau(\nabla_k u_i) (\nabla_i u_k)$. It was demonstrated in  \cite{fouxonlee} that Eq.~(\ref{flow}) holds also when gravity is included with a different formula for $\bm v(t, \bm x)$ in terms of $\bm u(t, \bm x)$ though. The condition of validity of the flow description of Eq.~(\ref{lw}) is that $St\ll 1$, independently of the strength of gravity, or $Fr\ll 1$ independently of $St$. Here, the dimensionless Froude number $Fr$ is defined as the ratio of typical acceleration of the Kolmogorov scale eddies $\epsilon^{3/4}\nu^{-1/4}$ and $\bm g$. Thus particles with large inertia, $St\gg 1$ can still form a smooth flow in space because of the smoothing action of gravity \cite{fouxonlee}. It is found that in all cases where the flow description holds, that flow is necessarily weakly compressible. 

Our study below is done for particles whose motion in space can be described by smooth weakly compressible flow as given by Eq.~(\ref{flow}). Thus it holds provided $\min[Fr, St]\ll 1$. The results do not depend on the detailed form of $\bm v(t, \bm x)$ due to the universality described in \cite{fouxon1}. We stress that the use of Eq.~(\ref{flow}) does not demand having approximately continuum distribution of particles in space. This equation can be used even when there is only one particle in the whole space. 

We describe the implications and limitations of the previous studies of Eq.~(\ref{flow}) for the finite number of particles in the volume where the discrete nature of particles is relevant. We use the theory constructed in \cite{fouxon1} based on the continuity equation for the spatial concentration of particles $n(t, \bm x)$. If we distribute the particles uniformly in space then their trajectories are attracted by the random multifractal set in space that has zero volume. After few Kolmogorov times $\sqrt{\nu/\epsilon}$ the trajectory that starts at some arbitrary point in space will reach a certain location on the definite multifractal set in space. This set, which can be called random attractor, will continuously change in time but its statistics are time-independent. Thus if we have only one particle in the flow volume then it will be somewhere on the attractor but we cannot know where. A few particles will also be randomly distributed over that attractor. In the limit of infinite number of particles, that are considered in the point particle approximation given by Eqs.~(\ref{lw}) $\&$ (\ref{flow}), the particles will cover the attractor continuously. 

The continuum theory constructed in \cite{fouxon1} demonstrates that statistics of the number of particles $N_l(\bm x)$ inside a ball of radius $l$ centered at $\bm x$ is log-normal. This theory holds in the continuum limit of infinite space-averaged concentration $\langle n\rangle$ - given by the total number of particles in the volume of the flow divided by the volume. Using the results of \cite{fouxon1} we find in terms of coarse-grained concentration $n_l=N_l/[(4\pi l^3/3)]$ that, 
\begin{align}
\lim_{\langle n\rangle\to\infty}\frac{\langle N_l^k\rangle}{\langle N_l\rangle^k}\!=\!\left(\frac{\eta}{l}\right)^{k(k-1)D_{KY}},\ \ \sigma^2\!=\!2D_{KY}\ln \left(\frac{\eta}{l}\right) \label{continuum},\\
P(N_l)\!=\!\frac{1}{N_l\sqrt{2\pi \sigma^2}}\exp\left(-\frac{\left[\ln \left(N_l/\langle N_l\rangle\right)\!+\!\sigma^2/2\right]^2}{2\sigma^2}\right),\nonumber
\end{align}
where angular brackets stand for spatial averaging and $\langle N_l\rangle=4\pi l^3 \langle n\rangle/3$. Here, $D_{KY}$ is the Kaplan-Yorke co-dimension of the random attractor \cite{KY} that because of the flow's weak compressibility is given by $D_{KY}=|\sum \lambda_i/\lambda_3|$. The Lyapunov exponents $\lambda_i$ give the rates of growth of infinitesimal lines, surfaces and volumes of the particles $\lambda_1$, $\lambda_1+\lambda_2$, $\sum_{i=1}^3\lambda_i$, respectively. It is demonstrated in \cite{fouxon1} based on the formula of \cite{itzhakgrisha} for $\sum\lambda_i$ that 
\begin{eqnarray}&&
D_{KY}=\frac{1}{2|\lambda_3|_i}\int_{-\infty}^{\infty} \langle \nabla\cdot\bm v(0)\nabla\cdot\bm v(t)\rangle_i dt,
\label{eq:dky}
\end{eqnarray}
where the trajectories $\bm q_i(t, \bm x)$ 
\begin{eqnarray}&&
\partial_t\bm q_i(t, \bm x)=\bm v_i[t, \bm q_i(t, \bm x)],\ \ \bm q_i(t=0, \bm x)=\bm x,
\end{eqnarray}
that define $\lambda_3$ and the correlation function 
\begin{equation}
\langle \nabla\cdot\bm v(0)\nabla\cdot\bm v(t)\rangle=\int \frac{d\bm x}{V}\nabla\cdot\bm v(0, \bm x)\nabla\cdot\bm v[t, \bm q_i(t, \bm x)],
\label{eq:sumlambda_i}
\end{equation} are those of the solenoidal (incompressible) component $\bm v_i$ of $\bm v$. The case where gravity is negligible and $St\ll 1$ results in $\bm v=\bm u-\tau\left[\partial_t\bm u+(\bm u\cdot\nabla)\bm u\right]$, so that we have $\bm v_i\approx \bm u$. For the incompressible component of the particles' flow one can use the underlying turbulent flow $\bm u$. Then $\lambda_3$ is the third Lyapunov exponent of the fluid particles, $\bm q_i(t, \bm x)$ are Lagrangian trajectories labeled by their initial positions and $\langle \nabla\cdot\bm v(0)\nabla\cdot\bm v(t)\rangle$ is the ordinary different time correlation function of $\nabla\cdot\bm v=-\tau(\nabla_k u_i) (\nabla_i u_k)$ (we observe that the Navier-Stokes equations give $-(\nabla_k u_i) (\nabla_i u_k)=\nabla^2 p$ where $p$ is the turbulent pressure). However, in the case of $Fr\ll 1$ the incompressible component can differ from $\bm u$ so that for instance $(\lambda_3)_i$ is very different from the third Lyapunov exponent of the fluid particles. It can be written in terms of the energy spectrum of turbulence \cite{fouxonlee}. 

\section{Statistics of discrete particles}\label{sec:theory}

\begin{figure}
\includegraphics[width=8.7 cm,clip=]{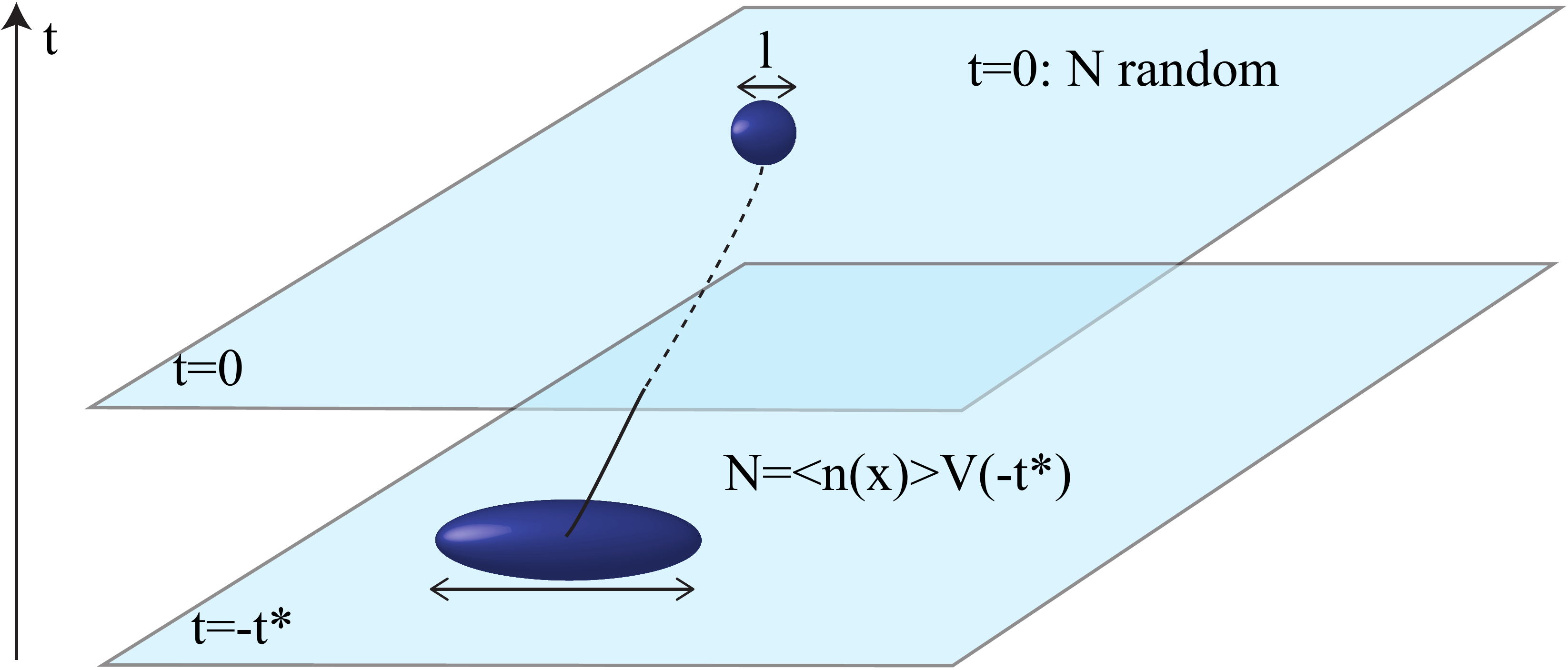}
 \caption{The event of finding $k$ particles in the ball of radius $l$ is a composition of two occurrences: incompressible turbulence randomly brings particles inside a certain volume $V(-t^*)$ with largest size $\eta$ that in time $t^*$ will be transformed in the considered ball. Compressibility of the flow of particles is relevant only for the volume compression stage. The probability of bringing $k$ particles in $V(-t^*)$ is the corresponding probability for the Poisson process because incompressible turbulence distributes particles over the volume of the flow uniformly and independently. The volume $V(-t^*)$ is however random which produces Eq.\,\ref{ppr}.}   
\label{fig2} \end{figure}

In this section, we consider how the discreteness of particles changes the statistics of the number of particles $N_l(0, \bm x)$ that at $t=0$ are located inside a ball of radius $l\ll \eta$, centered at $\bm x$ in comparison with the continuum theory. The theory for the finite number of discrete particles must reproduce the continuum theory in the limit of the infinite number of particles. We use the line of consideration of \cite{fouxon1} illustrated in Fig. \ref{fig2}. This is based on the conservation of the (finite) number of particles inside the volume, which at $t=0$ is the considered ball of radius $l$. This conservation holds because the particles' motion is described by a smooth differentiable flow in space as described by Eq.~(\ref{flow}). We find the statistics by tracing the particles back in time to the moment $-t^*$ when the particles were independent (in fact this consideration is quite similar to that done for the Boltzmann equation where the particles before the collision are considered independent \cite{ll9}). This is based on the fact that clustering is a small-scale phenomenon, so the weak compressibility produces non-negligible corrections to the uniform spatial concentration that would hold for incompressible flow only at the smallest scales. The particles that are located at $t=0$ inside the considered ball came from larger separations in the past, where they moved independently until the short time interval near $t=0$ when they approached each other to a distance smaller than $\eta$ and started moving in the common velocity gradient (correlated over $\eta$). Here the independence of motion above $\eta$ must be understood in the sense that there is no formation of correlations of concentration - the particles' motion itself is correlated in the inertial range. We remark that the considerations below can be made more rigorous using separation scales which are much smaller than $\eta$ but over which the particles can be already considered independent (see \cite{fouxon1}). For clarity, we use only the scales $l$ and $\eta$ without changing the conclusions. 
 
We track the volume occupied by trajectories $\bm q(t, \bm x')$ back in time 
\begin{eqnarray}&&
\partial_t\bm q(t, \bm x')=\bm v[t, \bm q(t, \bm x')],\ \ \bm q(t=0, \bm x')=\bm x',
\end{eqnarray}
with $|\bm x'-\bm x|<l$. Only a finite number of trajectories $\bm q(t, \bm x')$ corresponds to actual particles. These are trajectories $\bm q(t, \bm x_i)$, where $\bm x_i$ are positions of particles inside the considered ball at $t=0$. The rest of the trajectories $\bm q(t, \bm x')$ are mathematical constructs. The volume occupied by $\bm q(t, \bm x')$ is transformed by the flow in an ellipsoid whose largest axis grows as $l\exp[|\lambda_3 t|]$, where $\lambda_3$ is the third Lyapunov exponent. This reaches $\eta$ at time $-t^*$ where $t^*=|\lambda_3|^{-1} \ln(\eta/l)$. Since over this scale the correlations of particles are negligible because of their motion in the common flow \cite{fouxon1} (the particles that move independently at scales larger than $\eta$, are particles brought randomly togehter by the flow below $\eta$), the number of particles inside the ellipsoid obeys the Poisson distribution. Here, we consider the number of particles inside the ellipsoid which is determined by the particles' motion at times smaller than $-t^*$ as independent of the ellipsoid's volume. This is because the volume is determined by the flow divergence over the time interval $(-t^*, 0)$, see Eq.~(\ref{v}) below, where the vicinity of $-t^*$ of the order of the divergence correlation time can be neglected. This neglect is possible because weak compressibility implies a small volume change over that time.

The Poisson distribution of the number of particles is fixed uniquely by the intensity (or the average) which is given by $\langle n\rangle V(-t^*)$ where $V(-t^*)$ is the volume of the ellipsoid. Since $V(-t^*)$ is random, we conclude that the distribution of particles is a Poisson distribution with random intensity. We have, 
\begin{equation} P[N_l=k]=\left\langle\frac{\left[\langle n\rangle V(-t^*)\right]^k\exp\left[-\langle n\rangle V(-t^*)\right]}{k!}\right\rangle_{V(-t^*)}\label{ppr}
\end{equation}
where the remaining averaging is over the statistics of $V(-t^*)$. This type of distribution is known as doubly stochastic Poisson or Cox processes \cite{grandell,kingman2}. Furthermore as the statistics of $V(-t^*)$ is log-normal, see \cite{fouxon1} and below, then the process is a log Gaussian Cox distribution introduced in \cite{moeller}.

We now consider the evolution of the volume $V(t)$ occupied by the trajectories $\bm q(t, \bm x')$ at times larger than $-t^*$. Since the largest size of the volume is smaller than $\eta$, we can use the equation for the infinitesimal volume evolution \cite{Batchelorbook}
\begin{eqnarray}&&
\frac{d\ln V}{dt}=\nabla\cdot \bm v[t, \bm q(t, \bm x)],\ \ V(0)=\frac{4\pi l^3}{3}, \label{v}
\end{eqnarray}
where $\nabla\cdot \bm v[t, \bm q(t, \bm x')]\approx \nabla\cdot \bm v[t, \bm q(t, \bm x)]$ for considered $\bm x'$ obeying $|\bm x'-\bm x|<l$. The solution gives,
\begin{eqnarray}&&
V(-t^*)=\frac{4\pi l^3}{3}\left(\frac{\eta}{l}\right)^{-\rho(\bm x)}.\label{ev}
\end{eqnarray}
where we introduced a Gaussian variable,
\begin{eqnarray}
&& \!\!\!\!\!\!\!\!\!\!\rho(\bm x)=\frac{1}{\ln(\eta/l)}\int_{-t^*}^0 \nabla\cdot \bm v[t, \bm q(t, \bm x)]dt .
\label{gaussrho}
\end{eqnarray}
The gaussianity of $\rho$ can be seen by writing the integral as sum of the integrals over disjoint intervals whose duration is the correlation time $\tau_{cor}$ of $\nabla\cdot \bm v[t, \bm q(t, \bm x)]$. Since the number of disjoint intervals is large, $t^*\gg \tau_{cor}$ and the integrals over the disjoint intervals are independent random variables then gaussianity follows from the central limit theorem. A rigorous proof can be constructed using the cumulant expansion theorem \cite{Ma}. The Gaussian distribution of $\rho$ is determined by the average and the dispersion \cite{fouxon1},
\begin{eqnarray}
&& \langle \rho\rangle=D_{KY},\ \ \langle \rho^2\rangle-\langle \rho\rangle^2=\frac{2D_{KY}}{\ln(\eta/l)}. \label{gs}
\end{eqnarray}
We find that 
\begin{equation} P[N_l=k]=\frac{\langle N_l\rangle^k}{k!}\left\langle
\left(\frac{\eta}{l}\right)^{-k\rho}\exp\left[-\langle N_l\rangle \left(\frac{\eta}{l}\right)^{-\rho}\right]\right\rangle_{\rho},\label{pdf}
\end{equation}
where $\langle N_l\rangle=\langle n\rangle (4\pi l^3/3)$. This form of $\langle N_l\rangle$ is necessary because 
\begin{eqnarray}&& \!\!\!\!\!\!\!\!\!\!\!\!\!\langle N_l\rangle=\int \frac{d\bm x}{V}N_{l}(\bm x)=\frac{N_{tot}(4\pi l^3/3)}{V},\end{eqnarray} where $N_{tot}$ is the total number of particles in the volume $V$ of the flow.

We verify that $\langle N_l\rangle=\langle n\rangle (4\pi l^3/3)$ directly from the definitions. Using that the moments of $N_l$ can be found by the two-step procedure: first making Poissonian average at fixed $V(-t^*)$ or $\rho$ and then averaging over $\rho$. Based on Eq.~(\ref{ev}), we find
\begin{eqnarray}&& \!\!\!\!\!\!\!\!\!\!\!\!\! \langle N_l\rangle=\langle n\rangle \frac{4\pi l^3}{3}\left\langle \left(\frac{\eta}{l}\right)^{-\rho}\right\rangle_{\rho}.\label{first} \end{eqnarray} Applying Eq.~(\ref{gs}) for an arbitrary Gaussian random variable $r$ we have $\langle\exp[r]\rangle=\exp\left[\langle r\rangle+\left\langle (r-\langle r\rangle)^2\right\rangle/2\right]$ we find readily
\begin{eqnarray}&& \!\!\!\!\!\!\!\!\!\!\!\!\!  \left\langle\left(\frac{\eta}{l}\right)^{-k\rho}\right\rangle_{\rho}=\left(\frac{\eta}{l}\right)^{k(k-1)D_{KY}}, \label{rbk}\end{eqnarray} for arbitrary $k$. Using the result for $k=1$ in Eq.~(\ref{first}) we confirm $\langle N_l\rangle=\langle n\rangle (4\pi l^3/3)$.

\subsection{Moments of the number of particles in a given volume}

We consider the low moments of $N_l$.  Since dispersion of the Poisson distribution is equal to the average then,
\begin{eqnarray}&& \!\!\!\!\!\!\!\!\!\!\!\!\!
\langle N_l^2\rangle=\langle N_l\rangle^2 \left\langle\left(\frac{\eta}{l}\right)^{-2\rho}\right\rangle_{\rho}+\langle N_l\rangle.\label{eq:poisson}
\end{eqnarray}
We find using Eq.~(\ref{rbk}) that,
\begin{eqnarray}&& \!\!\!\!\!\!\!\!\!\!\!\!\! \frac{\langle N_l^2\rangle}{\langle n\rangle^2[4\pi l^3/3]^2}=\left(\frac{\eta}{l}\right)^{2D_{KY}} +\frac{1}{\langle n\rangle [4\pi l^3/3]},\ \ l\ll \eta.
\label{eq:cluster_powerlaw}
\end{eqnarray}
The last term on the RHS illustrates the correction to the continuum approximation described by Eq.~(\ref{continuum}) due to the discreteness of the number of particles. That term disappears in the continuum limit of $\langle n\rangle\to\infty$. Since the first term on the RHS is larger than one, the continuum approximation is valid in the limit where typically the considered ball of radius $l$ contains a large number of particles, hence $\langle N_l(\bm x)\rangle\gg 1$. We can then write Eq.\,(\ref{eq:cluster_powerlaw}) differently, 
\begin{eqnarray}&& \!\!\!\!\!\!\!\!\!\!\!\!\!
\langle N_l^2\rangle=\langle N_l\rangle^2 \left(\frac{\eta}{l}\right)^{2D_{KY}}+\langle N_l\rangle.\label{eq:2ndmoment}
\end{eqnarray}
In this form the growth of deviations from Poissonicity at small scales becomes obvious. When the scale $l$ is not too small so that $(\eta/l)^{2D_{KY}}\approx 1$ (note that $2D_{KY}\ll 1$ so these scales can be much smaller than $\eta$) we have $\langle N_l^2\rangle=\langle N_l\rangle^2+\langle N_l\rangle$ - that is we recover the dispersion of the Poisson distribution. In contrast at smaller scales with $(\eta/l)^{2D_{KY}}>1$, deviations from Poissonicity occur that grow indefinitely as $l\to 0$. These deviations are the discrete counterpart of the continuum description of preferential concentration by the increase of the radial distribution function at small separations. 

Similarly, we consider the discreteness correction to the third moment. Using the third moment of the Poisson distribution we obtain that
\begin{eqnarray}&& \!\!\!\!\!\!\!\!\!\!\!\!\!
\langle N_l^3\rangle\!=\!\langle n\rangle^3 \langle V^3(-t^*)\rangle\!+\!3\langle n\rangle^2 \langle V^2(-t^*)\rangle+\langle N_l\rangle.\end{eqnarray} This results in,
\begin{eqnarray}&& \!\!\!\!\!\!\!\!\!\!\!\!\! \frac{\langle N_l^3\rangle}{\langle N_l\rangle^3}\!=\! \left\langle\left(\frac{\eta}{l}\right)^{-3\rho}\right\rangle_{\rho}+\frac{3}{\langle N_l\rangle}\left\langle\left(\frac{\eta}{l}\right)^{-2\rho}\right\rangle_{\rho}+\!\frac{1}{\langle N_l\rangle^2}. \end{eqnarray}
Using Eq.~(\ref{rbk}) we obtain,
\begin{eqnarray}&& \!\!\!\!\!\!\!\!\!\!\!\!\! \frac{\langle N_l^3\rangle}{\langle N_l\rangle^3}\!=\!\left(\frac{\eta}{l}\right)^{6D_{KY}}
\!\!+\!\frac{3}{\langle N_l\rangle}\left(\frac{\eta}{l}\right)^{2D_{KY}}\!+\!\frac{1}{\langle N_l\rangle^2}.\label{third}\end{eqnarray}
If the average number of particles in the ball is large $\langle N_l\rangle\gg 1$, then the last two terms of Eq.\,(\ref{third}) that describe the corrections due to discreteness are negligible reproducing the continuum result. At the smallest scales there are strong deviations from Poissonicity that disappear at scales with $(\eta/l)^{6D_{KY}}\approx 1$. 

The consideration of higher-order moments can be performed based on,
\begin{eqnarray}&& \!\!\!\!\!\!\!\!\!\!\!\!\!
\exp[-\lambda]\sum_{k=0}^{\infty}\frac{k^n\lambda^k}{k!}=\sum_{k=1}^n \lambda^k S(n, k),
\end{eqnarray}
where $S(n, k)$ are Stirling numbers of the second kind \cite{weiss}. We find
\begin{eqnarray}&& \!\!\!\!\!\!\!\!\!\!\!\!\!
\langle N_l^n\rangle\!=\!\sum_{k=1}^n \langle N_l\rangle^k\left\langle\left(\frac{\eta}{l}\right)^{-k\rho}\right\rangle_{\rho} S(n, k),
\end{eqnarray}
which results in ($S(n, n)=S(n, 1)=1$),
\begin{multline}
\frac{\langle N_l^n\rangle}{\langle N_l\rangle^n}=\!\left(\frac{\eta}{l}\right)^{n(n-1)D_{KY}}+\sum_{k=2}^{n-1} \frac{S(n, k)}{\langle N_l\rangle^{n-k}}\left(\frac{\eta}{l}\right)^{k(k-1)D_{KY}} \\ + \frac{1}{\langle N_l\rangle^{n-1}},
\label{eq:higherordermoments}
\end{multline}
where the first term on the RHS is the continuum result and the rest of the terms on the RHS are the corrections due to discreteness. The considered cases of $n=2$, $3$ can be reproduced using the known values of $S(n, k)$.

\subsection{Voiding effect}

We start the study of the probability density function of $N_l(\bm x)$ from the probability that there are no particles in the volume of interest.  From Eq.~(\ref{pdf}) we get
\begin{eqnarray}&& \!\!\!\!\!\!\!\!\!\!\!\!\! 
P[N_l=0]=\left\langle\exp\left[-\frac{4\pi l^3\langle n\rangle}{3} \left(\frac{\eta}{l}\right)^{-\rho}\right]\right\rangle_{\rho},\label{prob}
\end{eqnarray}
where we used $\langle N_l\rangle=4\pi l^3\langle n\rangle/3$. In the limit of tracers, $D_{KY}\to 0$, the variable $(\eta/l)^{-\rho}$ does not fluctuate and equals its average which is one, so the equation reduces to the Poisson distribution.  Using $\langle (\eta/l)^{-\rho}\rangle=1$ and Jensen's inequality - $\langle \exp[X]\rangle> \exp[\langle X\rangle]$ - that holds for arbitrary non-constant random variable $X$ we find, 
\begin{eqnarray}&& \!\!\!\!\!\!\!\!\!\!\!\!\! 
P[N_l=0]>\exp\left[-\frac{4\pi l^3\langle n\rangle}{3} \right].\label{inequality}
\end{eqnarray}
Thus the void probability is larger than that for the Poisson point process (given by the RHS of Eq.\,(\ref{inequality})). This manifests clustering of particles in turbulence, implying larger probability of voids. 

Starting from Eq.~(\ref{prob}) and using Gaussianity of $\rho$, we conclude that the probability $P[N_l=0]$ of the void of size $l$ is the Laplace transform of the log-normal distribution.  In our case we have,
\begin{eqnarray}
&& \!\!\!\!\!\!\!\!\!\!P(N_l=0)=\sqrt{\frac{\ln(\eta/l)}{4\pi D_{KY}}}\int d\rho \exp\biggl[-\langle N_l\rangle \left(\frac{l}{\eta}\right)^{\rho}
\nonumber\\&&\!\!\!\!\!\!\!\!\!\!
-\frac{\left(\rho-D_{KY}\right)^2}{4D_{KY}}\ln\left(\frac{\eta}{l}\right)\biggr];\ \ \ \ \ \  \ l\ll\eta,\label{integral}  \end{eqnarray}
which in the limit of tracers (inertia vanishes), $D_{KY}\to 0$, reduces to the Poissonian $\exp[-\langle N_l\rangle]$. This limit however is slowly convergent because of the exponential $(l/\eta)^{\rho}$ in the exponent which has further implications that are considered below. We rewrite this formula in the standard form using the integration variable $y=(D_{KY}-\rho)\ln(\eta/l)$, 
\begin{eqnarray}
&& \!\!\!\!\!\!\!\!\!\!P(N_l=0)=F\left(\theta, \sigma^2\right),\ \ \theta=\langle N_l\rangle \left(\frac{l}{\eta}\right)^{D_{KY}}, \label{sigma}
\end{eqnarray}
where $\sigma^2$ is defined in Eq.~(\ref{continuum}) and we introduced the Laplace transform of the log-normal distribution with zero mean,
\begin{equation}
F(\theta, \sigma^2)=\sqrt{\frac{1}{2\pi \sigma^2}}\int_{-\infty}^{\infty} dy \exp\biggl[-\theta \exp\left(y\right)
-\frac{y^2}{2\sigma^2}\biggr]\label{lapl}.\end{equation}
We can infer from this the PDF $p(l)$ of the void radius $l$ that provides more direct information on void sizes. This PDF is defined so that $p(l)dl$ is the probability that the void centered at some $\bm x$ has a radius between $l$ and $l+dl$. We observe that the probability that $N_l=0$ coincides with the probability that the void has a size which is not smaller than $l$, hence $P(N_l=0)=\int_l^{\infty} p(l')dl'$. Differentiating this identity results in,   
\begin{multline}
p(l)=-\frac{dP(N_l=0)}{dl}=-\frac{(3+D_{KY})\theta\nabla_{\theta}F\left(\theta, \sigma^2\right)}{l}\\
+\frac{2D_{KY}\nabla_{\sigma^2}F\left(\theta, \sigma^2\right)}{l}, \label{deriv}\end{multline}
where we used a differentiation over $l$ at other parameters fixed. For instance, writing $\theta=\langle N_{\eta}\rangle (l/\eta)^{3+D_{KY}}$ we find $d\theta/dl=(3+D_{KY})\theta/l$. It is readily seen that the normalization condition $\int_0^{\infty}p(l)dl=1$ holds. It will be clear from the study of asymptotic forms of $P(N_l=0)$ below, that this formula implies a log-normal decay of $p(l)$ that is much slower than the exponential decay of the Poisson distribution. \\
\\
\emph{Average void size.}
We compute the average void size based on Eq.\,(\ref{deriv}), 
\begin{eqnarray}&& \!\!\!\!\!\!\!\!\!\!\!\!\! 
\langle l\rangle=\int_0^{\infty} lp(l)dl=\int_0^{\infty} P(N_l=0)dl. \label{eq:l}
\end{eqnarray}
Substituting Eq.\,(\ref{integral}) into Eq.\,(\ref{eq:l}) yields,
\begin{eqnarray}
&& \!\!\!\!\!\!\!\!\!\!\langle l\rangle=\int_0^{\eta} dl \int d\rho\sqrt{\frac{\ln(\eta/l)}{4\pi D_{KY}}}  \exp\biggl[-\langle N_{\eta}\rangle \left(\frac{l}{\eta}\right)^{3+\rho}
\nonumber\\&&\!\!\!\!\!\!\!\!\!\!
-\frac{\left(\rho-D_{KY}\right)^2}{4D_{KY}}\ln\left(\frac{\eta}{l}\right)\biggr];\   \label{avrlength}\end{eqnarray}
where we used $\langle N_{\eta}\rangle=4\pi \eta^3\langle n\rangle/3$. The integral is cut at $\eta$ because we use $P(N_l=0)$ for $l<\eta$. Equation (\ref{avrlength}) holds provided the self-consistency condition $\langle l\rangle\ll \eta$ is obeyed. We find \begin{eqnarray}
&& \!\!\!\!\!\!\!\!\!\!\frac{\langle l\rangle}{\eta}=\int_0^{1} dy \int d\rho\sqrt{\frac{\ln (1/y)}{4\pi D_{KY}}}  \exp\biggl[-\langle N_{\eta}\rangle y^{3+\rho}
\nonumber\\&&\!\!\!\!\!\!\!\!\!\!-\frac{\left(\rho-D_{KY}\right)^2}{4D_{KY}}\ln(1/y)\biggr]. \end{eqnarray}
The maximum of the exponent is taken at $y=0$ and $\rho=D_{KY}$ and given by zero. The case of interest is $\langle N_{\eta}\rangle\gg 1$ because otherwise the typical size of the void is of order $\eta$ or larger and does not belong to the domain of inertial effects considered here (this is not so at $D_{KY}\sim 1$ which is outside of the domain of our consideration). Then the width of integration over $y$ is $\langle N_{\eta}\rangle^{-1/(3+\rho)}\ll 1$. Then the integral over $\rho$ is strongly peaked at $\rho=D_{KY}$ and we can set in the integral $y^{3+\rho}\approx y^{3+D_{KY}}$. Subsequently, we can integrate over $\rho$ which results in, 
\begin{multline}
\frac{\langle l\rangle}{\eta}=\int_0^{1} dy  \exp\biggl[-\langle N_{\eta}\rangle y^{3+D_{KY}}\biggr]\approx \langle N_{\eta}\rangle^{-1/(3+D_{KY})}
\\
\Gamma\left[1+\frac{1}{3+D_{KY}}\right]. 
\end{multline}
Using that $D_{KY}\ll 1$ we obtain, 
\begin{eqnarray}
&& \!\!\!\!\!\!\!\!\!\! \langle l\rangle\approx \eta\langle N_{\eta}\rangle^{-1/3+D_{KY}/9}\Gamma(4/3)=l_P\langle N_{\eta}\rangle^{D_{KY}/9}, \label{respons}\end{eqnarray}
where we introduce the average void size of the Poisson point process, 
\begin{eqnarray}
&& \!\!\!\!\!\!\!\!\!\! l_P=(4\pi\langle n\rangle/3)^{-1/3}\Gamma(4/3). \end{eqnarray}The self-consistency of the performed calculation demands that the width of the integration over $\rho$ is much smaller than $D_{KY}$. Since the characteristic value of $y$ is $\langle N_{\eta}\rangle^{-1/(3+D_{KY})}$, then the condition is $D_{KY}\ln \langle N_{\eta}\rangle/3\gg 1$, where we keep $1/3$ to clearly point out that in $\langle N_{\eta}\rangle^{D_{KY}/9}=\exp[D_{KY}\ln \langle N_{\eta}\rangle/9]$ the exponent must be large. When $D_{KY}\ln \langle N_{\eta}\rangle\lesssim 1$ our formula can be used by order of magnitude predicting that $\langle l\rangle\sim l_P$. We remark that the self-consistency of the calculations demands that the final answer must be independent of the fact that $\eta$ is defined only up to factor of order one. In fact if we multiplied $\eta$ by factor of order one, $\langle N_{\eta}\rangle\propto \eta^3$ would change at most by order of magnitude. This would change $\langle l \rangle$ by factor of $10^{D_{KY}/9}\approx 1$ that is produce no appreciable change.  

We conclude that turbulence can have a strong effect on the average void size however small inertia is. If we consider the limit where the distribution of particles is quite dense so that $\langle N_{\eta}\rangle$ is so large that $\langle N_{\eta}\rangle^{D_{KY}/9}$ is also large than the typical void size is larger than $\langle n\rangle^{-1/3}$. Similar consideration can be performed for other moments $\langle l^k\rangle$.\\
\\
\emph{Approximate solution of $P(N_l=0)$.} There is no closed form for $F(\theta, \sigma^2)$ though there are well-working approximations, see for instance \cite{danish}. We consider the case where the argument of the exponent has a sharp maximum which includes both continuum and tracer limits. The position $y^*$ of the maximum is determined by,
\begin{eqnarray}
&& \!\!\!\!\!\!\!\!\!\! y^*=-\theta\sigma^2 \exp\left(y^*\right),\ \ y^*=-W(\theta\sigma^2),
\end{eqnarray}
where $W$ is the Lambert function, defined as solution to the equation $W(x)\exp\left[W(x)\right]=x$. Introducing $h(y)=-\theta \exp\left(y\right)-y^2/[2\sigma^2]$ we have $h'\left(y^*\right)=0$ and,
\begin{eqnarray}
&& \!\!\!\!\!\!\!\!\!\! h\left(y^*\right)=-\frac{W^2(\theta\sigma^2)+2W(\theta\sigma^2)}{2\sigma^2}\\&&\!\!\!\!\!\!\!\!\!\! h''\left(y^*\right)=-\frac{\theta\sigma^2 \exp\left(y^*\right)+1}{\sigma^2}=-\frac{1+W(\theta\sigma^2)}{\sigma^2}. \end{eqnarray}
Thus in quadratic approximation we find, 
\begin{multline}
P(N_l=0)\approx \sqrt{\frac{1}{2\pi \sigma^2}}\exp\left[-\frac{W^2(\theta\sigma^2)+2W(\theta\sigma^2)}{2\sigma^2}\right]\\
\int_{-\infty}^{\infty} dy \exp\left[-\frac{\left[1+W(\theta\sigma^2)\right](y-y^*)^2}{2\sigma^2}\right],\label{quadr}  
\end{multline}
which by integration gives
\begin{equation}
P(N_l=0)\approx\frac{\exp\left(-\left[W^2(\theta\sigma^2)+2W(\theta\sigma^2)\right]/[2\sigma^2]\right)}{\sqrt{1+W(\theta\sigma^2)}}. \label{gauss} 
\end{equation}
This approximation is valid provided the width $\sigma^2/\left[1+W(\theta\sigma^2)\right]$ of the Gaussian maximum in Eq.~(\ref{quadr}) is much smaller than one. Since $W(x)>0$ at $x>0$ then the Gaussian approximation holds when $\sigma^2\ll 1$ independently of the rest of the parameters. 
\\

We now consider two separate cases, namely (i) no preferential concentration at scale $l$ where we find that turbulence can still significantly increase the probability of large voids and (ii) the regime of strong clustering where we find that the typical void size increases strongly.\\
\\
\emph{Case $\sigma^2\ll 1$.}
This is the case where there is no preferential concentration at scale $l$. This is because the pair-correlation function of concentration is $\exp[\sigma^2]\approx 1$ see Eq.~(\ref{continuum}). One could think then that inertia is negligible at this scale and Poissonian void probability holds. This is not the case as the study of Eq.~(\ref{gauss}) reveals. Poissonian distribution holds only provided that besides $\sigma^2\ll 1$, the inequality $\theta\sigma^2\ll 1$ holds as well. Then using that $W(x)\approx x$ at small $x$ we have,
\begin{multline}
P(N_l=0)=\exp\left[-\frac{4\pi l^3\langle n\rangle}{3}\right],\\\ln\left[\left(\frac{\eta}{l}\right)^{2D_{KY}}\right]\ll 1,\ \ \langle N_l\rangle\ln\left[\left(\frac{\eta}{l}\right)^{2D_{KY}}\right]\ll 1,  \label{conds}
\end{multline}
where we used that $\theta\approx \langle N_l\rangle$ at $\sigma^2\ll 1$. However, if $\theta\sigma^2\approx \langle N_l\rangle \sigma^2$ is of order $1$ deviations from Poissonicity occur. The void probability becomes much larger than Poissonian in the limiting case where the concentration of particles is quite high so that $\langle N_l\rangle \sigma^2\gg 1$ despite $\sigma^2\ll 1$. Using that $W(x)\approx \ln x-\ln\ln x$ at large $x$ and consider 
for clarity the stronger inequality $\ln[\langle N_l\rangle \sigma^2]\gg 1$ we find from Eq.~(\ref{gauss}) that,
\begin{multline}
P(N_l=0)\approx \frac{1}{\sqrt{\ln[\langle N_l\rangle \sigma^2]}}\exp\left(-\frac{\ln^2[\langle N_l\rangle \sigma^2]}{4D_{KY}\ln(\eta/l)}\right),\label{void} \\
\ln[\langle N_l\rangle \sigma^2]\gg 1,\ \ \frac{\ln[\langle N_l\rangle \sigma^2]}{\sigma^2}\gg 1,\ \ l\ll \eta, 
\end{multline} 
where the second condition in the last line is the condition of validity of the Gaussian approximation which is relevant in the case of strong clustering $\sigma^2\gg 1$ only (otherwise it is implied by the first condition). The case of $\sigma^2\gtrsim 1$ and $\sigma^2\ll 1$ is considered below separately. Written with these conditions the formula covers the case of strong clustering $\sigma^2\gtrsim 1$ also.

We find that the decay of the probability of a large void in the void's volume is much slower than the exponential decay of the Poisson distribution. For the case under consideration without preferential concentration $\sigma^2\ll 1$, the squared logarithm in the exponent of Eq.~(\ref{void}) has a large prefactor. Thus the probability of finding voids is small being given by the exponent of a large negative number. However it can still be much larger than the Poissonian probability - the ratio of probabilities is proportional to $\exp\left(\langle N_l\rangle-\ln^2[\langle N_l\rangle \sigma^2]/[4D_{KY}\ln(\eta/l)]\right)$. This becomes infinite in the continuum limit $\langle n\rangle\to \infty$ at other parameters fixed (the limit can be turned dimensionless by multiplying with $\eta^3$) when $\langle N_l\rangle\to \infty$. Thus turbulence strongly increases the probability of large voids (here large voids are defined as voids whose probability is much less than one because of their large size).

\begin{figure}[h]
\includegraphics[scale=0.43]{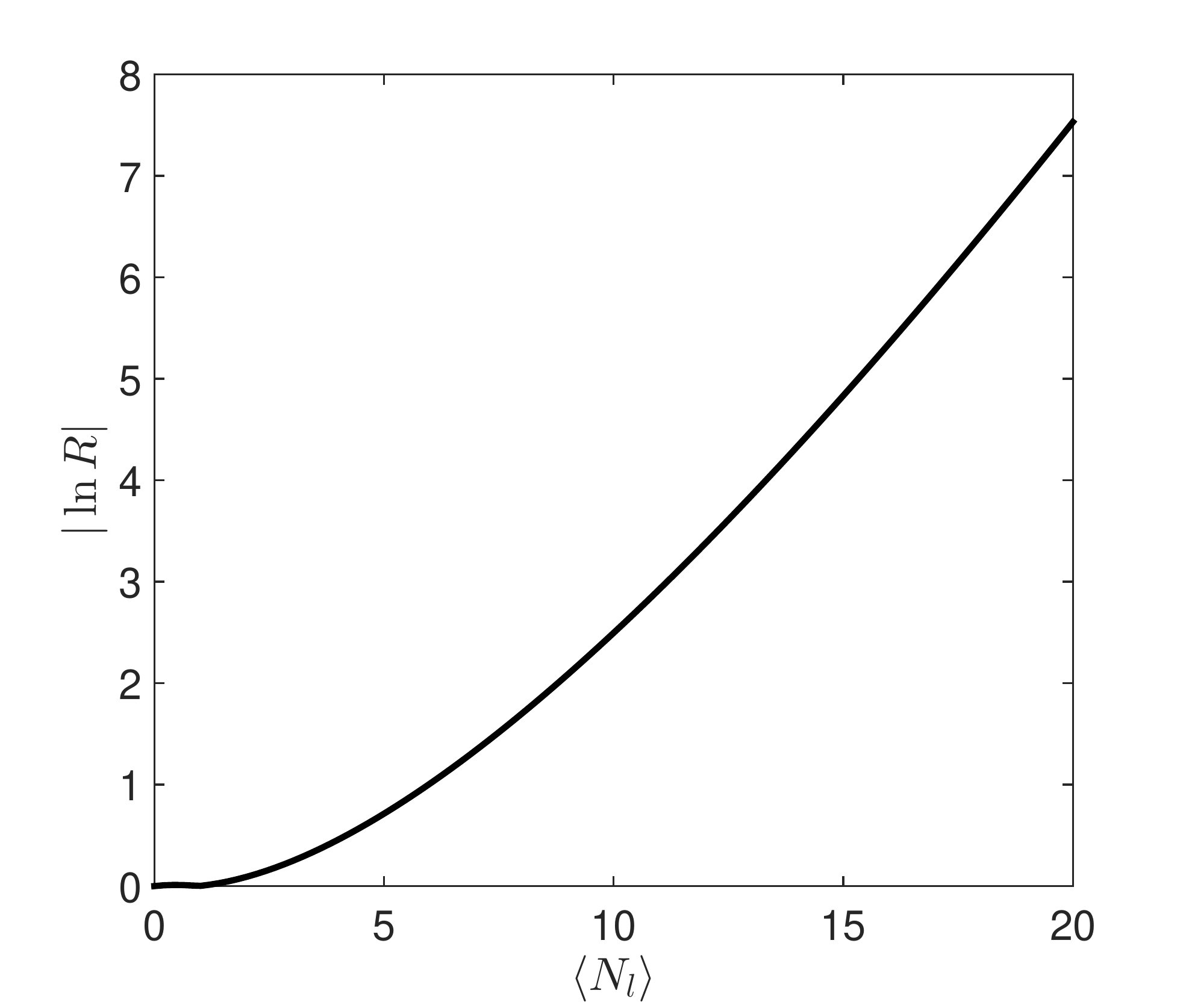}
\caption{Logarithm of ratio $R$ from Eq.\,\ref{eq:R} for the interval $\langle N_l\rangle=[0,20]$ using $\sigma^2=0.1$. The probability of finding voids becomes significantly larger than the one given by the Poisson distribution for values $\langle N_l\rangle\sigma^2\geq 1$. At larger $\langle N_l\rangle$ we observe a slow convergence towards linear behavior ($|\ln R|\approx \langle N_l\rangle$) which is reached at approximately $\langle N_l\rangle=10^4$.}
\label{fig:ratioR} 
\end{figure}

We considered the opposite limiting cases of $\theta \sigma^2\ll 1$ and $\ln[\theta\sigma^2]\gg 1$. In the limit of $\theta\sigma^2\ll 1$ the probability becomes Poissonian. When $\theta \sigma^2$ increases from small values at fixed $\sigma^2$ (for instance considering higher $\langle n\rangle$) the probability deviates from Poissonian probability necessarily becoming larger than that because of Eq.~(\ref{inequality}). We consider the deviation at arbitrary $\theta \sigma^2$ considering $\sigma^2\ll 1$ fixed so that Eq.~(\ref{gauss}) holds. We plot the logarithm of the ratio $R$, defined as $P(N_l=0)$ given by equation (\ref{gauss}) over the Poissonian probability,
\begin{eqnarray}&& \!\!\!\!\!\!\!\!\!\!\!\!\!\!\!
R(\langle N_l\rangle)=\frac{1}{\sqrt{1+W(\langle N_l\rangle\sigma^2)}}\nonumber\\&&\!\!\!\!\!\!\!\!\!\!\!\!\!\!\!\times\exp\left(\langle N_l\rangle-\frac{W^2(\langle N_l\rangle\sigma^2)+2W(\langle N_l\rangle\sigma^2)}{2\sigma^2}\right)\label{eq:R} \end{eqnarray} in Fig.\,\ref{fig:ratioR}, showing that larger values of $\langle N_l\rangle$ lead to a significant increase of $R$ and thus the probability of finding voids of particles.

Thus we derived the closed form of $P(N_l=0)$ at $\sigma^2\ll 1$ at arbitrary $\langle N\rangle$. We demonstrated that the void probability is significantly larger than Poissonian when $\langle N_l\rangle\sigma^2\gtrsim 1$.\\
\\
\emph{Case $\sigma^2\geq1$}. When there is preferential concentration at scale $l$ then the Gaussian approximation works provided $W(\theta\sigma^2)/\sigma^2\gg 1$.
This condition becomes true in the continuum limit where it becomes $\ln[\langle N_l\rangle ]/\sigma^2\gg 1$. This is how Eq.~(\ref{void}) works at $\sigma^2\gtrsim 1$. We remark that the Gaussian approximation becomes valid in the limit of large $\langle N_l\rangle$ independently of $\sigma^2$. 
The first of the conditions given by Eq.~(\ref{conds}) is identical to that for the second moment: if $(\eta/l)^{2D_{KY}}\approx 1$ then the clustering is insignificant and that moment becomes Poissonian. However for the third moment the criterion for Poissonicity is $(\eta/l)^{6D_{KY}}\approx 1$, see Eq.~(\ref{third}). The reason why in the second moment, a power-law appears in the criterion is the Chebichev inequality. If $(\eta/l)^{2D_{KY}}\approx 1$ then fluctuations of $N_l$ in the continuum approximation are small, $\langle N_l^2\rangle\approx \langle N_l\rangle^2$, see Eq.~(\ref{continuum}). Then the inequality implies that $(\eta/l)^{-\rho}$ in Eq.~(\ref{prob}) is weakly fluctuating, cf. Eq.~(\ref{eq:poisson}). Finally, since the average of $(\eta/l)^{-\rho}$ is one, then $(\eta/l)^{-\rho}\approx 1$ and Eq.~(\ref{prob}) reduces to the Poisson distribution. Though fluctuations of $(\eta/l)^{-\rho}$ are small, when the prefactor $\langle N_l\rangle$ in Eq.~(\ref{prob}) is very large these fluctuations are enhanced so much that they cannot be neglected and there are finite deviations from Poissonicity. This is the content of the second condition given in Eq.~(\ref{conds}). If this condition does not hold, the complete formula (\ref{gauss}) must be used. 

In the range of strong clustering $\sigma^2\gtrsim 1$ (that holds when $l$ gets smaller always) where $\sigma^2>1$ but $W(\theta\sigma^2)$ is not so large so that $\sigma^2/\left[1+W(\theta\sigma^2)\right]\gtrsim 1$ the Gaussian approximation given by Eq.~(\ref{gauss}) breaks down, cf. \cite{danish}. Despite considerable efforts, see \cite{danish,tellambura} and references therein, there seem to be no tractable formula for this case. However there are well-working numerical recipes \cite{danish,tellambura} which must be used for using our answer given by Eq.~(\ref{sigma}) in practice. 

\subsection{Probability of arbitrary number of particles}

Finally the probability of some arbitrary number $k$ of particles inside the ball of radius $l$ from Eq.~(\ref{pdf}) is, 
\begin{multline}
 P[N_l=k]=\frac{\langle N_l\rangle^k}{k!}
\sqrt{\frac{\ln(\eta/l)}{4\pi D_{KY}}}\int d\rho \exp\biggl[-\langle N_l\rangle \left(\frac{\eta}{l}\right)^{-\rho} \\
-\frac{\left(\rho-D_{KY}\right)^2+4D_{KY}k\rho}{4D_{KY}}\ln\left(\frac{\eta}{l}\right)\biggr].
\end{multline}
This obeys the normalization $\sum_{k=0}^{\infty}P[N_l=k]=1$. Completing to the square, $\left(\rho-D_{KY}\right)^2+4D_{KY}k\rho=\left(\rho+D_{KY}\left[2k-1\right]\right)^2-4kD_{KY}^2\left(k-1\right)$, we have,
\begin{eqnarray}&& \!\!\!\!\!\!\!\!\!\!\!\!\! P[N_l=k]=\frac{\langle N_l\rangle^k}{k!}\left(\frac{\eta}{l}\right)^{k(k-1)D_{KY}}
F\left(\theta_k, \sigma^2\right),\nonumber\\&&\!\!\!\!\!\!\!\!\!\!\theta_k=\langle N_l\rangle\left(\frac{\eta}{l}\right)^{D_{KY}[2k-1]},
\end{eqnarray}
where we used the integration variable $y=\ln(\eta/l)\left(D_{KY}\left[1-2k\right]-\rho\right)$ in combination with Eq.~(\ref{lapl}). This reduces to Eqs.~(\ref{sigma})-(\ref{lapl}) when $k=0$. We conclude that the PDF of the number of particles is determined by the Laplace transform of the log-normal distribution. 

In the continuum limit when $\langle n\rangle\to\infty$ at other parameters fixed, we have $\theta_k\to\infty$ and can use Gaussian approximation in finding $F(\theta_k, \sigma^2)$ which gives, 
\begin{multline} P[N_l=k]\approx \frac{\langle N_l\rangle^k}{k!}\left(\frac{\eta}{l}\right)^{k(k-1)D_{KY}}
\frac{1}{\sqrt{1+W(\theta_k\sigma^2)}}\\
\times \exp\left(-\frac{W^2(\theta_k\sigma^2)+2W(\theta_k\sigma^2)}{2\sigma^2}\right),
\end{multline}
cf. Eq.~(\ref{gauss}). We already proved that the continuum distribution is reproduced for the moments of $N_l$ so that the formula above must reduce to log-normal distributions which hold in the continuum limit as described by Eq.~(\ref{continuum}). It is possible to use the formula directly to reproduce the log-normal distribution as the leading order approximation for $\ln P[N_l=k]$. Higher-order terms in the expansion in logarithm are necessary (the leading order power in $k$ terms in $\ln P[N_l=k]$ vanish) but the corresponding formulas become cumbersome and are not presented here.  

\section{Radial distribution function in discrete description}\label{sec:rdf}

Due to the major importance of the radial distribution function (RDF) $g(\bm r)$ this Section demonstrates that discreteness of particles does not influence $g(\bm r)$. In fact the derivation of the RDF can be done when there are only two particles in the flow volume. We do not provide the details of the derivation that coincide with the continuum case studied previously see e.g. \cite{fouxon1}. 

The RDF counts the number of pairs of particles separated by distance $\bm r$,
\begin{eqnarray}&&\!\!\!\!\!\!\!\!\!\!\!\!\! 
g(\bm r)=\sum_{ik} \int \delta(\bm x_i-\bm x)\delta(\bm x_k-\bm x-\bm r)\frac{d\bm x}{\Omega},
\end{eqnarray}
where $\Omega$ is the total volume of the flow and the sum runs over all particles in the volume. The product of $\delta-$functions guarantees that only pairs separated by $\bm r$ are taken into account. For passing to the definition where $g(\bm r)$ is defined as fraction of the total number of pairs separated by $\bm r$, our definition is to be multiplied with the corresponding constant (we use the definition that is independent of the total number of particles which includes the case of two particles below). Using the microscopic definition of the concentration $n(t, \bm x)=\sum_i \delta(\bm x_i(t)-\bm x)$ we have $g(\bm r)=\langle n(\bm x)n(\bm x+\bm r)\rangle$ where angular brackets stand for spatial average over $\bm x$. In this way it is clear that the pair correlation of concentration obtained in the continuum limit gives $g(\bm r)$.

In the following we demonstrate that the steps in the derivation of the pair-correlation function \cite{fouxon1} can be performed without taking the continuum limit. We use the identity for the particle's trajectory $\bm x(t)$,
\begin{eqnarray}&&\!\!\!\!\!\!\!\!\!\!\!\!\! \delta(\bm x(0)-\bm x)=\delta[\bm x(t)-\bm q(t, \bm x)]\\&&\!\!\!\!\!\!\!\!\!\!\!\!\!\exp\left[-\int_t^0 \nabla\cdot\bm v[t', \bm q(t', \bm x)]dt'\right],\nonumber
\end{eqnarray}
obtained from
\begin{eqnarray}&&\!\!\!\!\!\!\!\!\!\!\!\!\! \det[\nabla_k f_i(\bm x)]\delta[\bm f(\bm x)]=\delta(\bm x-\bm x_0),\ \ \bm f(\bm x_0)=0,
\end{eqnarray}
using the function $\bm f(\bm x)=\bm q(t, \bm x)-\bm x(t)$. This identity is the discrete counterpart of the solution of the continuity equation for concentration. We can then repeat the steps in the derivation of the pair-correlation function in the language of discrete particles. If there are only two particles in the volume with trajectories $\bm x_1(t)$ and $\bm x_2(t)$ , we consider the joint probability density function of positions of two particles,
\begin{eqnarray}&&\!\!\!\!\!\!\!\!\!\!\!\!\!
P(\bm r)=\langle \delta(\bm x_1(0)-\bm x)\delta(\bm x_2(0)-\bm x-\bm r)\rangle.
\end{eqnarray}
We have
\begin{multline}
 P(\bm r)=\left\langle\delta(\bm x_1(t^*)-\bm q(t^*, \bm x))\delta(\bm x_2(t^*)-\bm q(t^*, \bm x+\bm r))
\right.\\
\left.\exp\left[\!-\!\int_{t^*}^0\!\ \!dt'\left(\nabla\!\cdot\!\bm v[t', \bm q(t', \bm x)]\!+\!\nabla\!\cdot\!\bm v[t', \bm q(t', \bm x\!+\!\bm r)]\right)\!\right]
\right\rangle.\nonumber \end{multline}
where $t^*=-|\lambda_3|^{-1}\ln(\eta/r)$ is the last time that trajectories separated by $\bm r$ at $t=0$ were separated by distance $\eta$. Using that the term in the last line is approximately independent of the rest of the terms we can perform independent averaging over the terms in the first and second line \cite{fouxon1}, 
\begin{multline} 
P(\bm r)=\left\langle\delta(\bm x_1(t^*)-\bm q(t^*, \bm x))\delta(\bm x_2(t^*)-\bm q(t^*, \bm x+\bm r))
\right\rangle\nonumber\\ 
\left\langle\!\exp\left[\!-\!\int_{t^*}^0\!\ \!dt'\left(\nabla\!\cdot\!\bm v[t', \bm q(t', \bm x)]\!+\!\nabla\!\cdot\!\bm v[t', \bm q(t', \bm x\!+\!\bm r)]\right)\!\right]
\!\right\rangle\!.\nonumber
\end{multline}
The average in the first line is $P(\eta)$ because at time $t^*$ the trajectories are separated by $\eta$. Since at scale $\eta$ the particles are independent then $P(\eta)=\Omega^{-2}$. The average in the last line is the same as that for the pair-correlation function of concentration in continuum theory. We find, 
\begin{eqnarray}&&\!\!\!\!\!\!\!\!\!\!\!\!\! 
P(\bm r)=\frac{1}{\Omega^2}\left(\frac{\eta}{r}\right)^{2D_{KY}},\ \ r\ll \eta. 
\end{eqnarray}
When more than $2$ particles are present, the contributions of different pairs are summed producing the result identical with that found in the continuum approximation. A similar result is valid for the joint probability density function of the position of $N$ particles.

Hence, we could demonstrate that the prediction of the power-law dependence of the RDF on the distance between the particles is identical in discrete and continuum descriptions. This is why the previous studies that often determined the RDF at scales where the average number of particles is small, so that the continuum approximation breaks down, could rely on results of the continuum theory \cite{collins}. 

\section{Numerical Verification}\label{sec:numerics}
\subsection{Numerical set-up}\label{subsec:setup}

In order to verify and analyze the theoretical results presented in the previous section (Sec.\,\ref{sec:theory}) we investigate the distribution of inertial particles in a chaotic Arnold-Beltrami-Childress ($ABC$) flow. This flow features a three-dimensional, incompressible velocity field and is an exact solution of the Euler equation. Previous studies have proven the generally chaotic features of this flow field \cite{arnold,dombre}. For the observation of the previously presented clustering phenomenon and its statistical description, a chaotic flow such as the $ABC$ flow is sufficient to illustrate the main effects. 

The $ABC$ flow guarantees that the incompressibility condition - $\bm\nabla\cdot \bm u=0$ - is precisely fulfilled at every point in the flow field. This is much more difficult to achieve in experiments or even in direct numerical simulations of turbulent flow where generally the flow is incompressible down to machine precision at collocation points but non-zero at arbitrary Lagrangian particle positions due to interpolation (and other possible) errors. Such artifacts might confound the effects under consideration here. In $ABC$ flow, the three components of the velocity vector $\bm u$ are determined by

\begin{equation}
\begin{gathered}
u_1\,=\,A\sin(x_3)+C\cos(x_2)\\
u_2\,=\,B\sin(x_1)+A\cos(x_3)\\
u_3\,=\,C\sin(x_2)+B\cos(x_1).\\
\label{eq:abcflow}
\end{gathered}
\end{equation}

Here, $A$, $B$ and $C$ define the space-independent amplitudes of the velocity components and $x_1,x_2$ and $x_3$ represent the three spatial coordinates of the system. 

In order to provide sufficiently chaotic advection of the particles and to avoid trapping particles in certain regions of the flow (KAM tori), the three amplitude parameters $A,B\,\&\,C$ are refreshed every $200$ computational time steps. In this way the amplitude parameters take on independent random values between $0$ and $1$. 

For the advection of particles we seed a total number of $10^8$ particles randomly in space at $t=0$ in a cubic volume with side length $2\pi$. Implemented periodic boundary conditions at all side walls of the cubic domain guarantee a constant number of particles in the domain at all times. 
The fluid particle velocity at each time step is determined by Eqs.\,(\ref{eq:abcflow}) at the particle location. The particles are advected in time using the forward Euler method  with a time step, $dt=0.005$, which provided sufficient accuracy. 

\begin{figure*}
\includegraphics[scale=0.4]{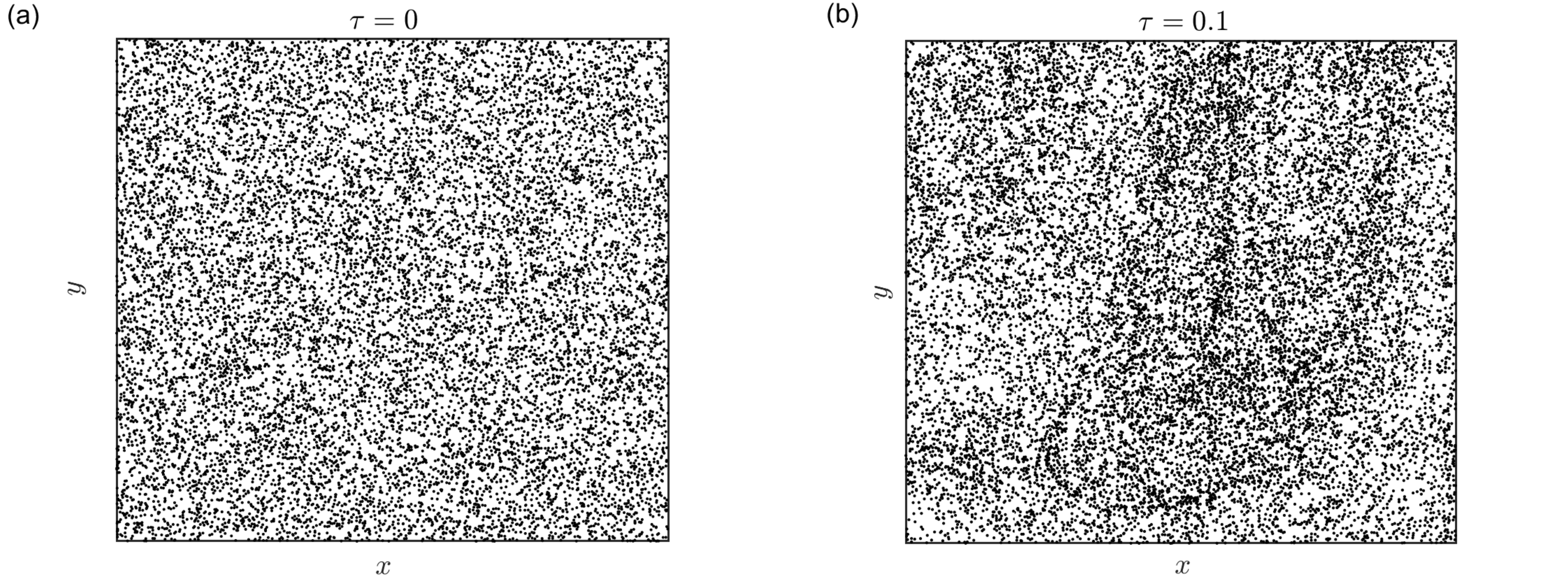}
\caption{Snapshot of  the particle distribution in a $\pi/10\times\pi/10$ window in the central slice of the investigation volume at $t=150$ for: (a) tracers ($\tau=0$), (b) inertial particles ($\tau=0.1$).}
\label{fig:Np_snapshots} 
\end{figure*}

Particles are advected for a total duration of $t=150$ for two cases: (i) passive tracers and (ii) inertial particles with particle response time $\tau=0.1$, which is the parameter that quantifies the strength of the particle inertia.
The tracer particle velocity is simply computed using Eqs.\,(\ref{eq:abcflow}). The inertial particle velocity $\bm v$ is obtained by adding an inertial drift to the fluid velocity $\bm u$, as described in Eq.\,(\ref{eq:particlevelocity}).

\subsection{Results}\label{subsec:results}

This section deals with the validation of the theory by numerical calculations of the inertial particle motion in chaotic $ABC$ flow. We analyze and compare the particle distribution using both the continuum theory and the corrections that take into account the discreteness of the number of particles in a given volume. We find that there is a need to include corrections when the average number of particles in the considered volume become of order $10$. Our results including the corrections are in excellent agreement with the predicted theory. 
Statistics are calculated for varying size of non overlapping cubic volumes within which $N_l$ is sampled. Samples are taken in the whole domain for a fixed time and the averaging is performed over space, which is expressed by $\langle . \rangle$. At $t=50-60$ the simulated passive particle ($\tau=0$)  distribution reaches statistical steady-state, while it took until $t=100$ for inertial particles ($\tau=0.1$). The results below are computed at $t=150$. To provide a qualitative impression of how tracer and inertial particles are distributed in space, a snapshot of the distribution of particles inside a small vertical window ($\pi/10\times\pi/10$) centered in the middle of the domain is illustrated in Fig.\,\ref{fig:Np_snapshots}. Despite the relatively small particle inertia, the difference between the tracer particle distribution (Fig.\,\ref{fig:Np_snapshots}a) and inertial particles (Fig.\,\ref{fig:Np_snapshots}b) is visible.
The tracer particles are randomly distributed. In contrast, the inertial particles are distributed randomly only at the initial time $t\,=\,0$, and later begin to concentrate in specific flow regions (see Fig.\,\ref{fig:Np_snapshots}b), thus causing the appearance of particle voids in other regions.

Using the theory of weakly compressible flow summarized in section \ref{sec:theory} above, allows to characterize the strength of the clustering by computing the Kaplan-Yorke co-dimension based on the Lyapunov exponents, i.e. $D_{KY}\,=\,\sum\lambda_i/\lambda_3$, doing this we find $D_{KY}=0.0122$ for the inertial particles. Thus, we verify that $\tau=0.1$ causes $D_{KY}\ll 1$, which is required for the proper application of the presented theory.

\begin{figure}[h]
\includegraphics[scale=0.43]{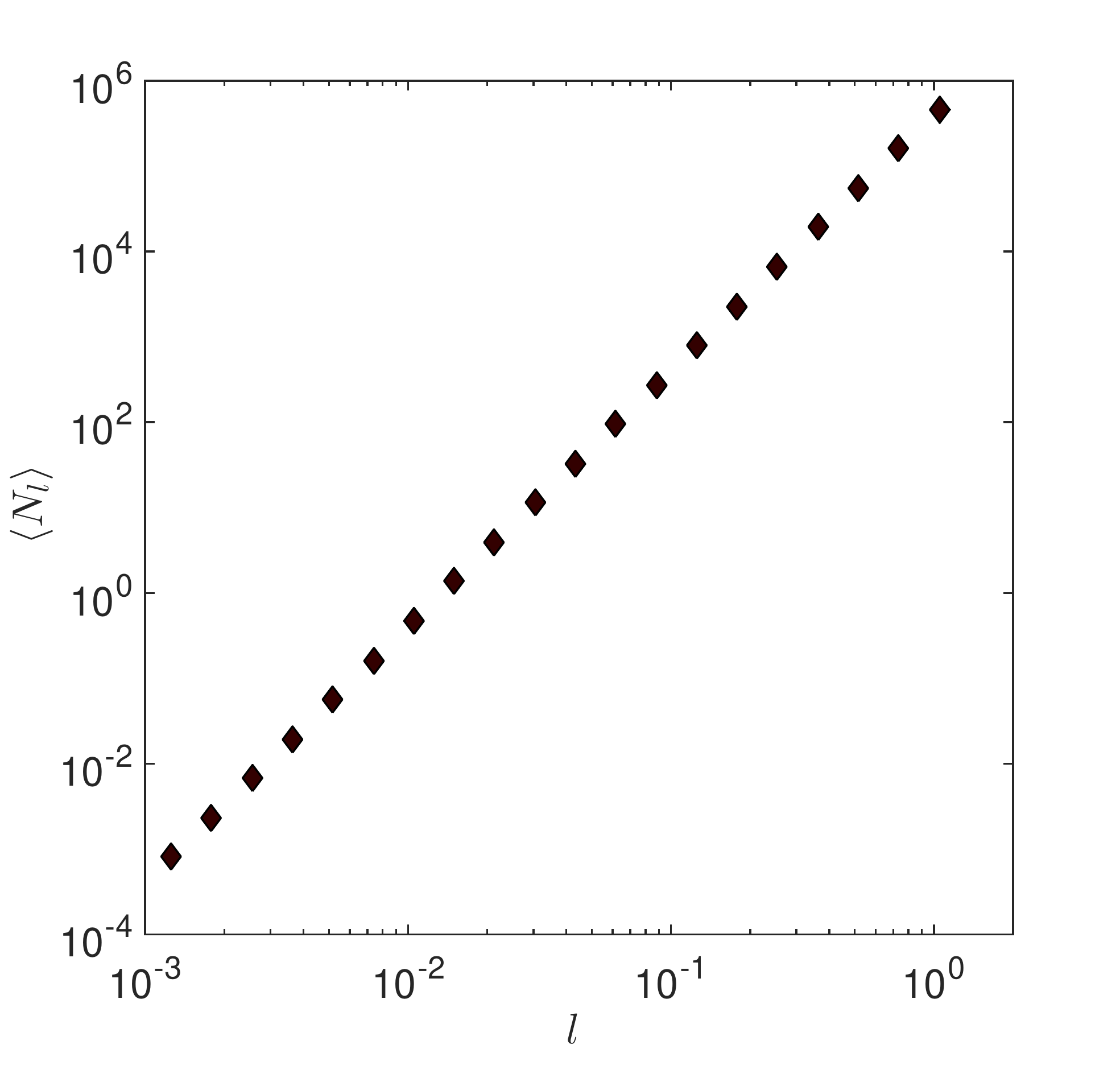}
\caption{Average number of particles found in a given cubic volume of size $l^3$.}
\label{fig:numofpart} 
\end{figure}

In the following we quantify the statistics of the number of particles in a given volume based on the presented theory. The correction due to discreteness of particles comes into play when the condition $\langle N_l\rangle\gg 1$ is not fulfilled anymore. We show the need for this correction on the examples of the second and third moment of $N_l$.

\begin{figure}
\includegraphics[scale=0.54]{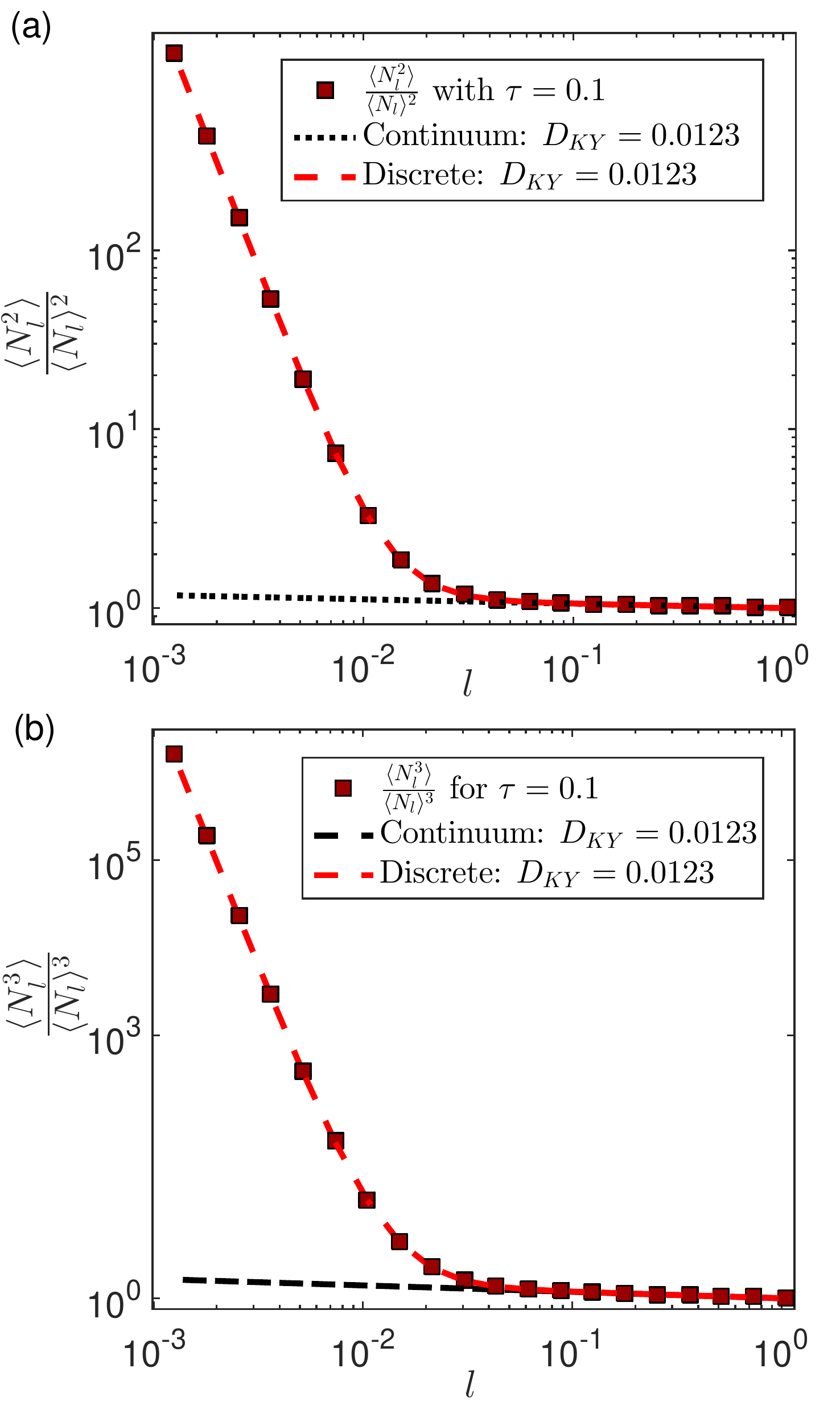}
\caption{(a): Second moment of $N_l$ for inertial particles (symbols). The continuum prediction $l^{-2D_{KY}}$ is the dotted black line. The red dashed line is the discrete fit of the form $l^{-2D_{KY}}+1/\langle N_l\rangle$. Both predictions use $D_{KY}=0.0123$. 
(b): Third moment of $N_l$ for inertial particles (symbols). The continuum prediction $l^{-6D_{KY}}$ is the dashed black line. The red dashed line is the discrete fit of the form $l^{-6D_{KY}}+3/\langle N_l\rangle l^{-2D_{KY}}+1/\langle N_l\rangle^2$. Both predictions use $D_{KY}=0.0123$.}
\label{fig:power-law_tua01}
\end{figure}

The average number of particles per bin $\langle N_l\rangle$ for varying bin size $l$ is illustrated in Fig.\,\ref{fig:numofpart}. This plot indicates that $N_l\gg 1$ might be valid in the region between $l\approx 0.1$ and the size of the domain. Hence, the discreteness correction will be necessary below that range.

\begin{figure}[h]
\includegraphics[scale=0.43]{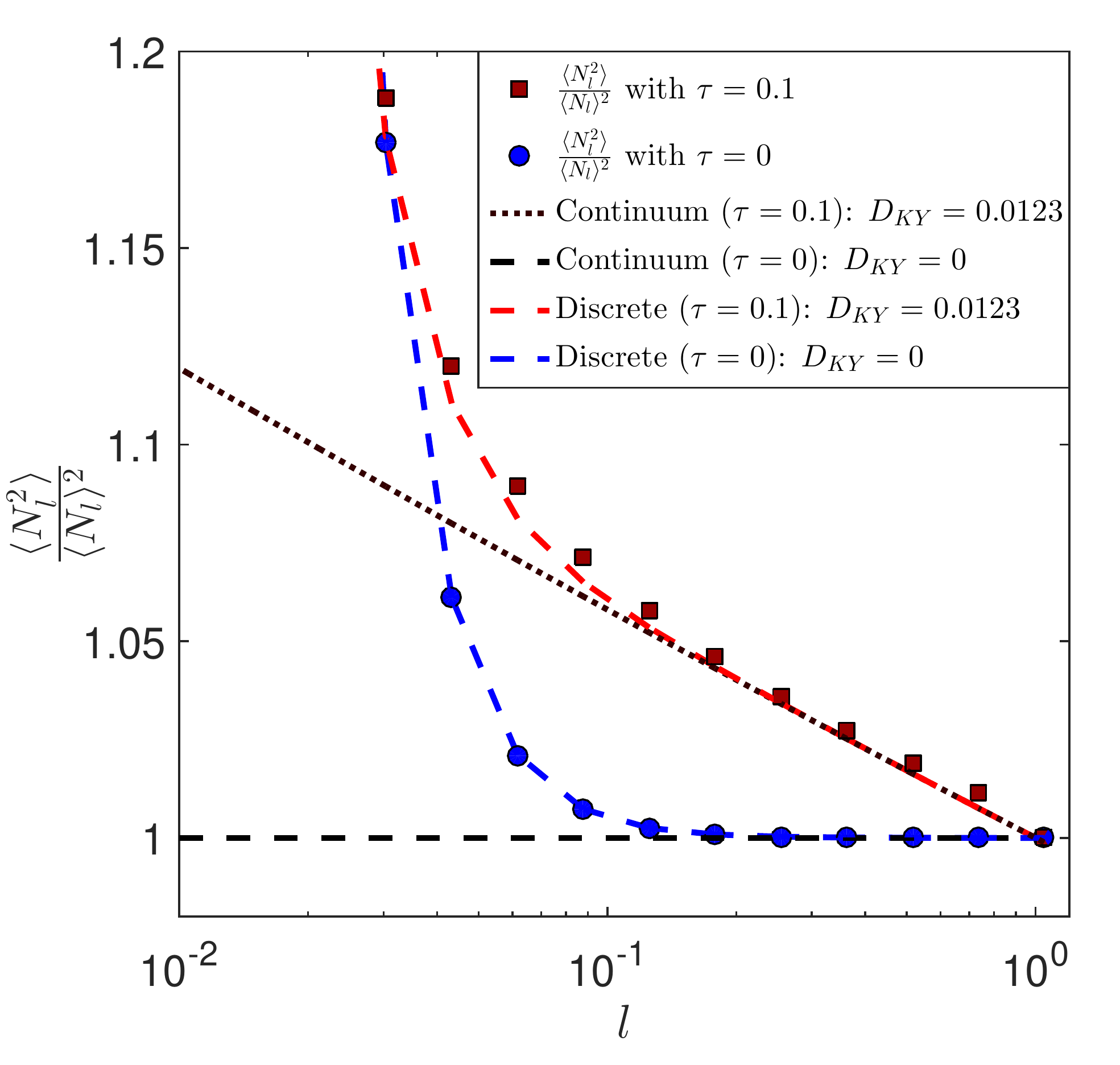}
\caption{Comparison of tracer and inertial particles for the second moment with a focus on the range l=[0.03,1]. The second moment is indicated by the smybols in red for inertial particles and the symbols in blue for tracer particles. The continuum prediction $l^{-2D_{KY}}$ is a dotted black line for inertial particles, where $D_{KY}=0.0123$ and a dashed black line for tracer particles (with $D_{KY}=0$). The discrete approach of the form $l^{-2D_{KY}}+1/\langle N_l\rangle$ for inertial particles is illustrated by the dashed red line ($D_{KY}=0.0123$) and as dashed blue line for tracer particles with $D_{KY}=0$.}
\label{fig:comp_tracevsinert} 
\end{figure}

Equation (\ref{continuum}) states that a simple power-law including $D_{KY}$ as exponent determines the statistics of the number of particles $N_l$ in the continuum limit. In cases where the discreteness of the number of particles matters, the moments of $\langle N_l\rangle$ are described by this power-law plus correcting terms, i.e. Eq.\,(\ref{eq:higherordermoments}) corresponding to Eq.\,(\ref{eq:2ndmoment}) for the second and Eq.\,(\ref{third}) for the third moment, respectively. Here, we can set $\eta=1$ since in $(\eta/l)^{2D_{KY}}$ we have $\eta$ of order one, which if it is raised in small power is approximately one, independent of its precise value. Figure \,\ref{fig:power-law_tua01} shows the difference between continuum and discrete approach on the example of the second and third moment for inertial particles. 
Symbols represent our data, while lines show the theory for the continuum and the discrete approach, respectively. In this flow the exponent $D_{KY}$ was fitted as $0.0123$ using the first decade $l=[0.1,1]$ in Fig.\,\ref{fig:power-law_tua01}a, where the average number of particles $\langle N_l\rangle$ is large enough for the continuum approach to be valid. This value is very close to the one found based on the Lyapunov exponents above, i.e. $D_{KY}=0.0122$. As illustrated in Fig.\,\ref{fig:power-law_tua01}, we get an excellent agreement between Eq.\,(\ref{eq:2ndmoment}) and our data, while the continuum approach prediction deviates from our data at $l<0.1$. A very similar result is obtained for the third moment shown in Fig.\,\ref{fig:power-law_tua01}b: there is excellent agreement between our data and the prediction for the discrete particle distribution (Eq.\,\ref{third}), while the continuum line deviates from our data for small $l$. It is seen that the deviation from the continuum prediction can be very strong for smaller $l$ (and $\langle N_l\rangle$ respectively), as anticipated in the theory section above.

Figure \,\ref{fig:comp_tracevsinert} shows the second moment of $\langle N_l\rangle$ for both passive and inertial particles from our data (symbols) and from continuum and discrete predictions (lines) for both kinds of particles. The passive tracer data is one for large enough $l$, where the continuum description applies, but starts deviating for $l<0.1$ and the trend perfectly matches the discrete prediction. The inertial particles show the power-law behavior discussed before and corrections apply roughly at the point where tracers deviate from $1$. Note that for very small $l$, corrections become even stronger for tracers than for inertial particles.

\section{Discussion and Conclusion}

We provided a detailed theoretical analysis of statistics of inertial particles transported by turbulence including discreteness of matter. The statistics of the number of particles in small volumes was found including the probability of finding voids. Previous studies did not take into account the discreteness of particles. We show that it can not be neglected in situations where the average number of particles in the considered volume is less than $10$, which will always occur in any real setting below a certain length scale. Since most experiments and numerical simulations deal with a limited number of particles the presented theory may be useful in numerous applications dealing with particle measurements in flows.

Our theory demonstrates that inertial particles in turbulent (or chaotic) flow distribute in space according to a Poisson process with log-normal random intensity. We derive corrections for the moments of the number of particles in small volumes due to particles' discreteness. The presented theory is based on an analysis of small-scale flow structures and can be applied to almost all turbulent and chaotic flows independent of their large-scale flow properties. We validate the theory through numerical experiments using a chaotic $ABC$ flow with periodic refreshment of amplitudes. There is very good agreement between our data and the theory as exemplified by the second and third moment of the number of particles.

We described the voiding effect of turbulence where the void probability is significantly increased in comparison with Poisson's void probability. Voiding consists of both increasing the typical size of the void and the probability of supervoids of size much larger than the typical one. 

We demonstrated that turbulence increases the typical (defined as average) void size $\langle n\rangle^{-1/3}$ of an ideal gas by a factor of $\langle N_{\eta}\rangle^{D_{KY}/9}$. We considered only the case where the average number of particles $\langle N_{\eta}\rangle$ at scale $\eta$ is much larger than one because otherwise the typical void size would be larger than $\eta$ and inertial effects would be irrelevant in the studied regime of $D_{KY}\ll 1$. The typical void size is as for independent particles $\langle N_{\eta}\rangle^{D_{KY}/9}\sim 1$. Due to the smallness of $D_{KY}$, this is a frequent occurrence. However at $\langle N_{\eta}\rangle^{D_{KY}/9}\gg 1$ there is a parametric increase of the typical void size, which is caused by turbulent vortices. In fact, however small $D_{KY}$ is, for very large $\langle N_{\eta}\rangle$ we will have a finite effect. For instance if $\langle N_{\eta}\rangle$ is of order of macroscopic numbers of $10^{22}$, we have an order of magnitude increase of the typical void size for $D_{KY}\sim 0.4$. Though this value of $D_{KY}$ is not very small it is plausible that the theoretical predictions would still work well. The limitation of realizability would come from the demand that the particles do not influence turbulence significantly and their hydrodynamic interactions can be neglected. However, there seems to be no a priori reason why this situation could not be feasible. 

The other limitation is that $\langle N_{\eta}\rangle$ cannot be made larger than $(\eta/a)^3$ by order of magnitude due to the finite size of the particles. Thus for water droplets in clouds the typical value of $\eta/a\sim 10^2$ would limit $\langle N_{\eta}\rangle$ by $10^6$ (in reality much smaller numbers hold since $\langle N_{\eta}\rangle\sim 1$ is quite typical). These numbers indicate that there would be no effect for $D_{KY}\ll 1$. 

The numerical factor of $1/9$ seems to be suppressing the increase of the typical void size quite strongly. Approaching an asymptotic continuation of the formula to $D_{KY}\sim 1$ (which holds in clouds at $Fr\approx 0.05$ \cite{fouxonlee}) we must use the complete formula $\langle N_{\eta}\rangle^{D_{KY}/(9+3D_{KY})}$ (in our calculations we neglected the $3D_{KY}$ factor in the denominator). If this formula is used with $D_{KY}=1$ then for getting an order of magnitude increase of the void size, $\langle N_{\eta}\rangle \sim 10^{12}$ would be required. However, small changes in this formula can easily deplete the necessary $\langle N_{\eta}\rangle$ by orders of magnitude. Thus we leave the question of finding the conditions where turbulence increases the typical void size by one order of magnitude for future work. 

Another facet of the voiding effect could be observable more readily. There is a strong increase of the probability of large voids whose size is much larger than the typical size. Remarkably, this effect can hold at scales $l$ where there is no preferential concentration so $D_{KY}\ln (\eta/l)\ll 1$. At those scales the pair-correlation function of the particle concentration $\langle n(0)n(\bm r)\rangle=\langle n\rangle^2(\eta/r)^{2D_{KY}}$ is that of independent particles or tracers, $\langle n(0)n(\bm r)\rangle\approx \langle n\rangle^2$. The radial distribution function (RDF) $g(\bm r)=\langle n(0)n(\bm r)\rangle/\langle n\rangle^2$ (we use a different normalization from the main text for clarity here) equals that of independent particles, $g(\bm r)\approx 1$. Thus the RDF of inertial particles is equivalent to the RDF of tracers or independent particles, hence one could think that inertia is irrelevant at those scales and Possonian void probability holds. This is not the case. We demonstrated that if the average number of particles inside the ball of the considered radius $\langle N_l\rangle=\langle n\rangle(4\pi l^3/3)$ is quite large $\langle N_l\rangle\gtrsim  [D_{KY}\ln (\eta/l)]^{-1}\gg 1$, then the probability of having large voids is significantly larger than for Poissonian statistics.

This shows that the rare event where a local turbulent vortex consistently pushes particles out of its core leads to a higher contribution than Poissonian chance-type formation of the void in the void probability. Thus the formation of voids is a stronger effect than preferential concentration of particles at scale $l$. Preferential concentration necessitates the formation of voids - the flux of particles in some regions produces voids in deserted regions. However, the converse is not true - holes can exist without clusters. Thus even at very small inertia (as measured by the dimensionless Stokes number $St$ when gravity is negligible) turbulence has a profound effect on the formation of voids of particles. 

We conclude that in considering the impact of turbulence on distributions of inertial particles, the study of the void probability can be instructive. This is reinforced by the fact that the void probability determines the statistics of the point distribution completely \cite{moller2}.
 
Though preferential concentration is always relevant at the smallest scales ($\langle n^2\rangle$ diverges) in practice it is usually relevant when $D_{KY}$ is not too small. For practical uses of the theory constructed under the assumption $D_{KY}\ll 1$ it is often necessary to continue the predictions asymptotically to the region of $D_{KY}\sim 1$. For instance when gravity is negligible and $St\ll 1$ the continuum theory predicts that the pair correlation function is a pure power law $(\eta/r)^{2D_{KY}}$ provided that $D_{KY}\ll 1$. In the case of $D_{KY}\sim 1$ the power-law still holds \cite{collins} empirically but with an exponent that coincides with the small $St$ prediction only by order of magnitude. By analogy, we consider it plausible that a Poisson distribution with random intensity applies at $D_{KY}\sim 1$ but with intensity statistics different from the case of $D_{KY}\ll 1$.

The study presented in this paper concentrates on the case where the particles' motion can be described by spatial flow. In the case where gravity is negligible ($Fr \gtrsim 1$) this confines the study to the case of $St\ll 1$. Considering the influence of increasing $St$ on the statistics of spatial distribution of particles we can concentrate on the study of the void probability. Thus a good modeling of the void probability can provide the key to the description of particle statistics. The immediate question raised is the role of the sling effect introduced in \cite{FFS} (and later in different nomenclature in \cite{caustics}) that was observed experimentally in \cite{Bewley} recently. In this phenomenon turbulent vortices of larger than average strength, vigorously throw particles out of their cores thus setting conditions for creating a void. Outside the vortices collisions happen faster than in the typical regions where vortices are calmer because different streams of particles intersect. The sling effect provides a significant contribution in the collision kernel of particles at moderately small Stokes number \cite{FFS}. Thus voids can be expected to be the next regions where multi-streaming occurs. In other words the number of voids of size of order $\eta$ can be expected to be similar to the number of slings which is quite well-studied \cite{pumirec}. Studies of the void probability at moderately small $St$ along this and other lines is future work. Once the probability of voids is found this opens the way for studies of long-term survival of interacting particles, where reactions make one of the particle type disappear. This type of particles survives for long times in void regions of the "predator" type of particles \cite{zeldovichovchinnikov}. Our results indicate that the formation of voids for inertial particles is more likely to happen than for tracers. Thus inertia of "predator" particles increases the long-time survival probability of reacting particles. Similar conclusions hold for other types of reactions including mating of living organisms. A quantitative study of concrete cases is future work.

\section{Acknowledgments}

We thank D. Krug for his help on Fig.\,\ref{fig2} and M. van Reeuwijk for his contribution on Lagrangian particle tracking algorithms. I. F. thanks E. Bodenschatz and H. Xu for providing preliminary data on in situ observations of voids of droplets in clouds. Financial support from the Swiss National Science Foundation (SNSF) under Grant No. 144645 is gratefully acknowledged.

\bibliography{biblio}

\end{document}